\documentclass[aps,prl,twocolumn,showpacs]{revtex4}
\usepackage{amsbsy,latexsym}
\usepackage{amsfonts}
\usepackage{amssymb}
\usepackage[mathscr]{eucal}
\def\H{{\sf H}}
\def\K{{\sf K}}
\newcommand{\dg}{{\rm d} g}
\newcommand{\al}{\alpha}
\newcommand{\be}{\beta}

\newcommand{\la}{\lambda}
\newcommand{\si}{\sigma}
\newcommand{\id}{\mathbb I}
\newcommand{\CF}{\boldsymbol n}
\newcommand{\aver}[1]{\langle #1 \rangle}
\newcommand{\Tr}{{\rm Tr}\,}
\newcommand{\M}{{\sf{M}}}

\newcommand{\ket}[1]{\vert #1 \rangle}
\newcommand{\bra}[1]{\langle #1 \vert}

\def\SU#1{\mathbb{SU}(#1)}

\begin{document}
\title{
Efficient use of quantum resources for the transmission of a reference frame} 
\author{G.~Chiribella}
\author{G.~M.~D'Ariano}
\altaffiliation[Also at ]{Center for Photonic Communication and
  Computing, Department of Electrical and Computer Engineering,
  Northwestern University, Evanston, IL  60208}
\author{P.~Perinotti}
\author{M.~F.~Sacchi}
\affiliation{QUIT Quantum Information
Theory Group of the INFM, unit\`a di Pavia}
\homepage{http://www.qubit.it}
\affiliation{Dipartimento di Fisica
``A. Volta'', via Bassi 6, I-27100 Pavia, Italy}

\begin{abstract}
We propose a covariant protocol for transmitting reference frames
encoded on $N$ spins, achieving sensitivity $N^{-2}$ without the need
of a pre-established reference frame and without using entanglement
between sender and receiver. The protocol exploits the use of
equivalent representations, which were overlooked in the previous
literature.
\end{abstract}

\maketitle In the ideal world of classical physics spatial directions
and reference frames can be communicated with arbitrary accuracy using
classical communication and a pre-established common frame, or by just
sending physical objects, such as gyroscopes. In the second case, if
Alice wants to send a frame to Bob, she needs only to align the
rotation axes of her gyroscopes with the directions she wants to
communicate, before sending them to Bob. Once Bob has measured the
direction of the gyroscopes, a common reference frame has been
established. Clearly, in the real world arbitrary accuracy is limited
by quantum fluctuations. However, similarly to the case of phase
estimation \cite{holevo,DMS}, we can learn how to harness the quantum
laws in order to achieve the ultimate precision limits of the communication
protocol.
\par The primitive systems that one can use for communication of
reference frames are quantum spins, since they can be considered as
elementary quantum gyroscopes. In this scenario, Alice transmits a
Cartesian reference frame by preparing $N$ spins in a quantum state
$\ket{A}$ which is related to her set of Cartesian axes
$\CF^{(A)}\doteq \{n^A_x, n^A_y, n^A_z\}$ and by sending them to
Bob. With respect to Bob's axes $\CF^{(B)}\doteq \{n^B_x, n^B_y, n^B_z
\}$, such a state corresponds to $\ket{A_g} \doteq
U^{\otimes N}_g \ket{A}$, where the unitary matrix $U_g$ represents
the rotation $g$ connecting Bob's frame to Alice's one, namely
$\CF^{(A)}=g~ \CF^{(B)}$. Now, Bob's task is to estimate the rotation
$g$ of the state $\ket{A_g}$, and then to align his axes with Alice's
frame. It is worth noting that such a scheme works without the need of any
pre-established reference frame. Notice also that the problem of aligning reference
frames using quantum spins is formally equivalent to the problem of estimating
unknown $\SU 2$ rotations (which is the same problem of estimating the
dynamics of an unknown qubit gate \cite{prl86,prl87,ajv}).

\par For the estimation of rotations with a finite number $N$ of spins
there is a nonzero probability of error which vanishes in the limit of
infinite $N$. Now the issue is to optimize the accuracy of the
estimation for a given $N$, by properly choosing Bob's measurement and
Alice's input state $\ket{A}$.  In the recent literature
\cite{frames-ps,frames-bbm,frames-ps2,frames-dc-bbm} much progress has
been made in this direction, and a number of strategies have been
proposed in specific cases. Nevertheless, in some of these works
\cite{frames-bbm,frames-ps2} it was argued that equivalent
representations of $\SU 2$ are redundant for encoding rotations, and
this oversight led to false claims of optimality in Ref.
\cite{frames-bbm}, where an asymptotic average error $1/N$ was found.
In this Letter, we show that, on the contrary, equivalent
representations play a crucial role in enhancing the sensitivity of
the estimation, since the inclusion of multiple equivalent
representations increases the dimension of the Hilbert space available
to storing information.  Moreover, we resolve a long-standing
controversy over whether the optimal strategy is covariant or not. In
Ref. \cite{frames-ps2} a non-covariant strategy is shown to do better
(with an error scaling as $1/N^2$) than the covariant strategy in Ref.
\cite{frames-bbm}. While the latter strategy was mistakenly thought to
be best, it appeared that the best covariant strategy was not optimal.
The present paper resolves the puzzle by showing that the optimal
covariant strategy does just as well as those presented in Ref.
\cite{frames-ps2} with an asymptotic error $1/N^2$.\par 

Finally, as we will show, there is a relation between the present
scheme and the entangled protocol of Ref. \cite{ajv}, with the role of
entanglement here played by equivalent representations.\par

Let us now summarize the main points in the problem of estimating
$\SU2$ rotations.  The most general estimation strategy that Bob can
perform---including both measurements and data analysis---is described
by a POVM, namely by a set of positive operators $\{M(g)\}$ in the
Hilbert space of N spins such that $\int \dg M(g)=\id$, with the
integral extended to the whole $\SU2$ group, and $\dg$ denoting the
invariant Haar measure on $\SU2$, normalized such that $\int \dg =1$.
The probability density of estimating $g$ when the true rotation is
$g_*$ is given by the Born rule: $p(g|g_*) \doteq \Tr[M(g)
\ket{A_{g_*}}\bra{A_{g_*}}]$. Finally, the efficiency of a strategy is
defined in terms of the transmission error
\begin{eqnarray}
e(g,g_*)\doteq\sum_{\al=x,y,z}|g n_{\al}^B-g_*n_{\al}^B|^2 \;,
\end{eqnarray}
which quantifies the deviation between the estimated axes and
the true ones. The maximization of the efficiency then corresponds to
the minimization of the average error
\begin{equation}\label{error}
\aver{e}=\int \dg_* \int \dg~ p(g|g_*) e(g,g_*)~.
\end{equation}
Notice that we have assumed a uniform \emph{a priori} distribution
$\dg_*$ for the true rotations, according to the fact that $g_*$ is
completely unknown. Since the function $e(g,g_*)$ enjoys the
invariance property $e(g,g_*)=e(hg,hg_*)$ for any $h \in \SU2$, as
proved by Holevo \cite{holevo} there is no loss of generality in
assuming that Bob's strategy is described by a \emph{covariant} POVM,
namely
\begin{equation}\label{covpovm} M(g) \doteq U_g^{\otimes N}~ \Xi~ U_g^{\dag~\otimes
N},
\end{equation} 
with $\Xi$ a positive operator. This fact relies on the covariance of
the set of input states. Indeed, for an arbitrary POVM $N(g)$ one can
always construct a covariant one with the same average error,
corresponding to $\Xi \doteq \int {\rm d}g~ U_g^{\dag~\otimes N} N(g)
U_g^{\otimes N}$.\par 

Let us now enter the core of our method.  In what follows, our aim
will be to use equivalent representations for constructing a highly
efficient reference state $\ket{A}$ in the space $\H^{\otimes N}$ of
$N$ spins. For this purpose, $\H^{\otimes N}$ can be conveniently
decomposed in terms of the Clebsch-Gordan series, i.e. as direct sum
of orthogonal subspaces which are irreducible under the action of
$\SU2$ rotations, namely
\begin{equation}\label{decomp}
\H^{\otimes N} = \bigoplus_{j=0(\frac{1}{2})}^{J} \bigoplus_{\alpha=1}^{n_j} \H_{j\alpha}~.
\end{equation}
Here $j$ represents, as usual, the quantum number of the total angular
momentum: it runs from $0$ ($\frac{1}{2}$) to $J = \frac{N}{2}$ for
$N$ even (odd), and labels the equivalence class of each irreducible
representation. On the other hand, $\alpha$ is a degeneracy index
labeling different equivalent representations in the same
class $j$.  For example, with three spins one has ${\bf
  \frac{1}{2}^{\otimes 3}}= {\bf \frac{3}{2}} \oplus {\bf \frac{1}{2}}
\oplus {\bf \frac{1}{2}}$, so that for the class $j= \frac{1}{2}$
there are two  equivalent irreducible representations 
corresponding to two orthogonal subspaces. The number
$n_j$ of equivalent representations in the class $j$ is given by \cite{cem}
\begin{equation}\label{multiplicity}
n_j={2j+1\over J+j+1}\pmatrix{2J\cr J+j}~.
\label{n_j}
\end{equation}
In each invariant subspace $\H_{j\alpha}$, we can introduce the basis
$\{\ket{j\alpha,~ m}\,;\ m=-j, \dots, j\}$ made of eigenvectors of
the $z$-component of the total angular momentum. With respect to these
bases, $\SU2$ rotations are represented by the ordinary Wigner
matrices $U^{(j)}_{nm}(g)$, namely 
\begin{equation}\label{matrix-rep}
U_g^{\otimes N}\ket{j\alpha,~m}=
\sum_{n=-j}^{j} U_{nm}^{(j)}(g)\ket{j\al,~n}~.
\end{equation}  Notice that two
vectors $\ket{j\alpha,~m}$ and $\ket{j\beta,~m}$ belonging to
different orthogonal subspaces $\H_{j\al}$ and $\H_{j\be}$ transform
in the same way under $\SU2$ rotations. Let's define then the
operator 
\begin{eqnarray}
T_{\al \be}^{(j)} \doteq \sum_{m=-j}^j
\ket{j\al,~m}\bra{j\be,~m} 
\;.
\end{eqnarray}
that takes a vector in the space $\H_{j\be}$ to the corresponding one
in $\H_{j\al}$. Using this operator we will compare vectors in
different equivalent subspaces, and we will say that two vectors
$\ket{\psi _{j \al}} \in \H _{j\al}$ and $\ket{\varphi _{j \be}} \in
\H _{j \be}$ are \emph{iso-orthogonal} if $\bra{\psi_{j\al}}
T^{(j)}_{\al\be} \ket{\varphi_{j\be}}=0$.

As opposite to the approach used in the previous works, here the state
$\ket{A}$ will be chosen in order to use as many equivalent
representations as possible. For this purpose, the crucial point is
that the maximum number of representations one can exploit in the
class $j$ is not $n_j$, but $k_j \doteq \min\{n_j,2j+1\}$,
corresponding to the fact that equivalent representations are useful
only when one takes iso-orthogonal vectors in different
representations. The proof of this statement has been derived in
\cite{MLpovms} and relies on the fact that for any given vector
$\ket{A}$, there is always a rearrangement of the decomposition
(\ref{decomp}) such that $\ket{A}$ has components on at most $k_j$
representations from the class $j$, and these components are all
iso-orthogonal to each other. Using (\ref{multiplicity}), it is easy
to see that $k_{j}=2j+1$ for $j <J$ and $k_{J}=n_{J}=1$. Keeping this
in mind, we make the following choice for Alice's reference vector
\begin{eqnarray}\label{alice}
|A \rangle =A_{J}|J,J \rangle + \sum _{j=0(\frac
  12)}^{J-1}\sum_{\alpha =1}^{2j+1} \frac{A_j}{\sqrt{2j+1}} |j\alpha,~
m(\alpha)\rangle~,
\end{eqnarray} 
where without loss of generality $A_j \geq 0$, and $m(\alpha)$ is an
injective function, namely $m(\alpha )\neq m(\alpha ')$ if $\alpha
\neq \alpha '$, according to the idea of taking an iso-orthogonal vector for
each equivalent representation.  Notice that the term for $j=J$, which has
multiplicity $n_J=1$, has been chosen arbitrarily with $m=J$. However,
as we will see in the following, its contribution is negligible in the
asymptotic limit of large $N$.

\par Now we need to specify which covariant POVM Bob must use to
extract the rotation $g$ from the state $\ket{A_g}$, namely we must
provide the operator $\Xi$ in Eq. (\ref{covpovm}). First we observe
that, since the vector $\ket{A}$ lies in the invariant subspace of
$\H^{\otimes N}$
\begin{equation}
\K= \H_J \oplus \bigoplus_{j=0(\frac{1}{2})}^{J-1} \bigoplus_{\al
=1}^{2j+1} \H_{j\al}~,
\end{equation} the probability distribution
\begin{eqnarray}
p(g|g_*)= \bra{A_{g_*}}~U_g^{\otimes N} \Xi U_g^{\dag ~ \otimes N}
~\ket{A_{g_*}} 
\;
\end{eqnarray}
depends only on the restriction $\xi \doteq P \Xi P$, where $P$ is the
projection on $\K$. Second, instead of optimizing Bob's POVM in order
to minimize the transmission error (\ref{error}), here we will take
the \emph{maximum likelihood POVM} \cite{MLpovms}, namely the POVM
which maximizes the peak $p(g_*|g_*)$ in the probability distribution
$p(g|g_*)$. For this POVM one simply has $\xi = \ket{B}\bra{B}$, where
\begin{eqnarray}\label{MLpovm} \ket{B}&=& \sqrt{2J+1} \ket{JJ} 
\nonumber \\ && + \sum_{j=0 (\frac 12)}^{J -1} \sum_{\alpha=1}^{2j+1}
\sqrt{2j+1} \ket{j\alpha,~ m(\alpha)}~.
\end{eqnarray}  
We stress that in the eigenstates of Eq. (\ref{MLpovm}), the $z$-component of the total angular
momentum is referred to Bob's axes, hence the transmission
protocol does not require a common reference frame (we remind that Alice's state $\ket{A}$ is seen as $\ket{A_g}=U_g^{\otimes N} \ket{A}$ in Bob's reference frame). \par  With the previous
settings, the problem of optimizing the coefficients $\{A_j\}$ in the state
$\ket{A}$  in order to minimize the transmission error becomes
straightforward. First, one can note
\cite{frames-bbm} that $e(g|g_*)= 6-2\chi(gg^{* -1})$, where
$\chi(g) \doteq \sum_{m=-1}^{1} U^{(1)}_{mm}(g)$ is the
character of the Wigner matrices for $j=1$.  Then, minimizing
the average error $\aver{e}$ is equivalent to maximizing the average
character
\begin{equation} \label{averchi}
\aver{\chi} \doteq \int \dg \, \chi(g)~ \, p(g|e)~,
\end{equation}  $e$ denoting the identical rotation. Notice that the integral
over $g_*$ in (\ref{error}) has been performed by exploiting the
invariance property of covariant POVM's, i.e.  $p(g|g_*)= p(hg|hg_*)
\,,\quad \forall h \in \SU2$.  Using the identity
\begin{eqnarray}
&&  \int 
\dg ~ U^{(1)}_{mm}(g)~~ U^{(j)}_{rs}(g)~U^{(l)*}_{ik}(g) \nonumber\\  
&&=\frac{1}{2l+1}~ \langle 1m~jr | li\rangle ~~ \langle lk |1m~js \rangle~, 
\end{eqnarray} 
where $\langle 1m~jr| li \rangle$ denote the Clebsch-Gordan coefficients, and performing the sums over equivalent representations, we obtain 
\begin{equation}\label{Mbraket}
\aver{\chi}=  \sum_{j,l=0(\frac 12)}^{J} A_j M_{jl} A_l \equiv  
A^{T} \mathsf{M} A~,
\end{equation}
where $A$ denotes 
the column vector $( A_J, A_{J-1}, \dots, A_{0(\frac
  12)})$, and $\mathsf{M}$ is the tridiagonal matrix
\def\xxx{\phantom{\displaystyle{.\over .}}}
\def\zzz{\hspace{.5em} }
\begin{equation}\label{matrix}
\mathsf{M}\doteq \pmatrix{
 {J\over J+1}     & {1\over \sqrt{2J+1}}              &         &   &  &  &\xxx\cr
 {1\over \sqrt{2J+1}}& 1                               &   1     &   & & \mbox{\LARGE0} &\xxx\cr
                   & 1 & 1 &\zzz 1                   &         &   &\xxx\cr
                   &   &\raisebox{1.3ex}[0ex][0ex]{1}&  \ddots & \ddots & &\xxx\cr
                   &   &                             & \ddots  &        & &\xxx\cr
 \hspace{.5cm} \raisebox{2.0ex}[1.5ex][0ex] {\LARGE0}\hspace{-.5cm} & & & &  1 \zzz & 1 &\zzz 1\xxx\cr
                   &   &   &  &  &    1  &   \zzz      \zeta\xxx } \ .
\end{equation}
Here $\zeta=0\ (1)$ for even (odd) values of $N$. Since the
normalization of Alice's vector implies $A^TA=1$, maximizing
$\aver{\chi}$ simply consists in finding the greatest eigenvalue $\la$
for the matrix $\mathsf{M}$: $\la$ is actually the maximum
$\aver{\chi}$ for our strategy and the optimal coefficients $\{A_j\}$
are the components of the corresponding normalized eigenvector.\par
For small $N$ one can easily perform numerical diagonalization: for
example with $N=3,\ 5,$ and $9$ one finds $\la= 1.3886,\ 2.0864$, and
$2.6294$, respectively. These values can be compared with those
obtained in Ref. \cite{frames-bbm} without the use of equivalent
representations: even for $N=3$ one can see a $17\%$ improvement of
$\aver{\chi}$. On the other hand, in the asymptotic limit of large $N$
an analytical treatment is possible, which is essentially based on the
fact that the contribution of the $J$ representation becomes
negligible. Let us denote the dependence on $N$ by writing
$\mathsf{M}^{(N)}$ and $\la^{(N)}$. If we introduce the matrix
$\mathsf{T}^{(N)}$ obtained from $\mathsf{M}^{(N)}$ by canceling the
first row and the first column (corresponding to ignore the $J$
representation) and call $\si^{(N)}$ its greatest eigenvalue, then we
have $\la^{(N)} \geq \si^{(N)}$. Nevertheless, it is also easy to see
that $\si^{(N+2)}\geq \la^{(N)}$, due to the fact that $0\leq
\mathsf{M}_{ij}^{(N)} \leq \mathsf{T}_{ij}^{(N+2)}$ for any $i,j$\
\cite{nota}. Hence, the asymptotic behavior of $\la^{(N)}$ is bounded
by $\si^{(N)}\leq \la^{(N)}\leq \si^{(N+2)}$. The matrix
$\mathsf{T}^{(N)}$ can be analytically diagonalized in terms of
Chebyshev polynomials, and its greatest eigenvalue is $\si^{(N)}= 1+2
\cos\left( \frac{2\pi}{N+1}\right)$. This implies the asymptotic
behavior $\aver{\chi} \sim 3-\frac{4\pi^2}{N^2}$, corresponding to the
following power law for the transmission error
\begin{equation} 
\aver{e} \sim \frac{8\pi^2}{N^2}~.
\end{equation}
Comparing this result with the behavior $\aver{e} \sim \frac{8}{N}$ of
\cite{frames-bbm}, one can observe a quadratic improvement due to the
use of equivalent representations.\par Notice that $\aver{e} \sim
\frac{8\pi^2}{N^2}$ is also the same efficiency of the protocol in
\cite{frames-dc-bbm}, where, by adopting the idea introduced in Ref.
\cite{ajv}, entanglement between sender and receiver is exploited, and
a collective measurement on two sets of $N$ spins is performed. With
respect to such protocol the present scheme provides a saving of
resources (i.e. half number of spins and no need of entanglement
between Alice and Bob), and, more important, does not require a
pre-established reference frame \cite{zero}.

\par There exists a connection between the present protocol  and
the entanglement-assisted one. In fact, let's introduce for any class
the \emph{representation space} $\H_j$ of dimension $2j+1$ and the
\emph{multiplicity space} $\M_j$ of dimension $n_j$, and write
$\ket{j\al,~ m}$ as $\ket{jm} \otimes \ket{\alpha} \in \H_j \otimes
\M_j$. Choosing $\{\ket{\al};\ \al=1, \dots,n_j\}$ as an orthonormal
basis for $\M_j$, one has 
\begin{equation}\label{isomorph}
\bigoplus_{\al =1}^{n_j}\H_{j\al} \equiv
\H _{j} \otimes \M_j~.
\end{equation}
By means of such isomorphism, we can rewrite our
choice of Alice's state as
\begin{equation}
\ket{A}\equiv A_J~ \ket{JJ}+ \sum_{j=0(\frac 12)}^{J-1} A_j~ \ket{E_j}~,
\end{equation}
where 
\begin{equation}\label{maxent}
\ket{E_j}\doteq \frac{1}{\sqrt{2j+1}} \sum_{\al =1}^{2j+1} \ket{jm(\al)}\otimes \ket{\al}       
\end{equation}
is a maximally entangled state between the representation space $\H_j$
and the multiplicity space $\M_j$ \cite{nota2}. If we neglect the $J$
term in $\ket{A}$, then we get a vector which is formally the same as
in \cite{frames-dc-bbm}. This means that the protocol exploiting
entanglement and $2N$ spins is reproduced using $N$ spins and without
entanglement between sender and receiver. We stress that here the
entanglement is between the representation and the multiplicity space
(which is not necessarily related to entanglement between the $N$
physical spins).

\par In conclusion, in this Letter we have shown how to exploit
equivalent representations of the rotation group for saving quantum
resources in transmitting a reference frame. A quadratic improvement
of the transmission efficiency has been achieved with respect to the
protocol of Ref. \cite{frames-bbm} which mistakenly neglects
equivalent representations. This is due to the fact that the use of
such representations provides more room for storing information. An
intuitive justification of this fact is provided by the maximum
likelihood strategy\cite{MLpovms}: in fact, the maximum likelihood
for a pure state is exactly proportional to the dimension of its orbit
under the action of the group, and for $N$ spins this is at most
$d_{max}= (2J+1) + \sum_{i=0(\frac{1}{2})}^{J-1} (2j+1)^2 \sim N^{3}$.
In our protocol this dimension is fully exploited by entangling the
representation space with the multiplicity space, whereas without such
entanglement one would obtain a dimension $d=(2J+1)
+\sum_{j=0(\frac{1}{2})}^{J-1} (2j+1)\sim N^2$. Notice that the use of
multiplicity spaces has been found to be necessary also in optimal
schemes for the transmission of elements of the permutation group
\cite{vkk}, and in achieving the optimal capacity for private
classical communication using a private shared reference frame
\cite{brs}.\par

Our results finally settle the controversy about covariance of the
optimal protocol, which was rised in Ref. \cite{frames-ps2}, by
providing a covariant scheme with the same performance $1/N^2$.\par

We also proved how the presence of equivalent representations provides
the remarkable possibility of reproducing the same efficiency of
covariant entangled protocols without the need of a pre-established
reference frame and without using entanglement between sender and
receiver. The present use of equivalent representations is a general
method which is not restricted to the transmission of reference
frames, and is expected to provide useful improvements also in other
estimation problems.

\par Discussions with R. Mu\~noz-Tapia about previous literature are
acknowledged. This work has been supported by INFM under PRA-2002-CLON
and by MIUR under Cofinanziamento 2003.


\begin{thebibliography}{99}
\bibitem{nota} We compare $N$ with
  $N+2$ terms, such that the considered matrices have the same value
  of the border term $\zeta$.
\bibitem{nota2} Notice that the property
  of iso-orthogonality becomes in the decomposition (\ref{isomorph}) 
  the customary orthogonality in the representation space
  $\H_j$.
\end{thebibliography}
\end{document}